\documentclass[aps,prb,twocolumn,epsf,floatfix]{revtex4}

\usepackage{natbib}

\usepackage{graphicx}

\usepackage{amssymb}

\begin{document}



\title{The Effect of Local Structure and Non-uniformity on 
Decoherence-Free States of Charge Qubits}


\author{Tetsufumi Tanamoto and Shinobu Fujita}

\affiliation{Advanced LSI Technology Laboratory, Toshiba Corporation, Kawasaki 
212-8582, Japan}

\begin{abstract}
We analyze robustness of decoherence-free (DF) subspace
in charge qubits when there are a local structure and non-uniformity 
that violate collective decoherence measurement condition.
We solve master equations of up to four charge qubits and a detector
as two serially coupled quantum point contacts (QPC) with an island 
structure. We show that robustness of DF states is strongly affected 
by local structure as well as by non-uniformities of qubits. 
\end{abstract}
\maketitle


\section{Introduction}
Decoherence-free(DF) states \cite{Zanardi} are useful for collective 
decoherence environment, even if there is a small symmetry breaking 
perturbation parameterized by  $\eta$ in the order of O($\eta$) 
($\eta \!\ll\!1$)\cite{Bacon}. 
Nowadays, experiments for DF states have been successful
up to four qubits in photon system\cite{Bourennane}
and in nuclear magnetic resonance (NMR)\cite{Viola}. 
However, in solid-state qubits, it seems to be more difficult to 
realize the collective decoherence environment than in the case of
optical or NMR qubits.
This is because we could not prepare plenty of qubits with 
mathematically exact size, because the sizes of 
Cooper-pair box\cite{Nakamura,Schoelkopf} 
or quantum dot (QD)\cite{Tarucha,Wiel,Hayashi,Gorman} 
are less than hundreds of nm, 
and moreover, localized trap sites generated in the fabrication process disturb 
the collective decoherence environment. 

We theoretically describe the effect of non-uniform environment 
on DF states of charge qubits composed of coupled QDs, 
considering the measurement process, by using time-dependent 
density matrix (DM) equations.
The charge distribution of the qubits changes the QPC current 
capacitively, resulting in detection and 
corruption of charged state (backaction)\cite{tana1}.
Thus, measurement is a basic and important decoherence process 
that should be investigated in detail.
In Ref.\cite{tana2}, we discussed the robustness of DF states 
under non-uniformity of qubits when they are detected by 
a simple structureless QPC detector.
However, generally speaking, the solid-state qubit system is 
arranged compactly to 
avoid extra noises, and 
therefore, qubits have a tendency to be often affected 
by geometrical local structures such as electrodes or electrical wires. 
In addition, when there are local structures and large non-uniformities, 
it is possible that the four-qubit DF states 
are less robust than non-DF states.
In such cases, for example, 
two qubit non-DF state and the singlet state 
would be appropriate to constitute two logical states 
instead of using four-qubit DF states, 
because defects and fault rates will increase as
the number of qubits increases. 

\begin{figure}
\begin{center}
\includegraphics[width=6.5cm]{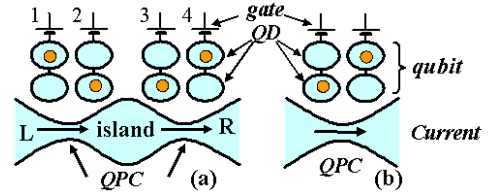}
\end{center}
\caption{Qubits that use double dot charged states are capacitively 
coupled to a QPC detector. 
}
\label{QPC}
\end{figure}

In this paper, we discuss the effect of local structures 
on the DF states combining with the non-uniformity of qubits 
based on the setup shown in Fig.~\ref{QPC}(a) and 
compare them with two-qubit states shown in Fig.~\ref{QPC}(b). 
We study the robustness of four-qubit DF states 
written as:
\begin{eqnarray}
|\Psi_1^{[4]}\rangle_{(1234)}\! &\!=\!&\!
 2^{-1}(|01\rangle\!-\!|10\rangle)_{(12)}
 \otimes (|01\rangle\!-\!|10\rangle)_{(34)},\ \ \ \ \nonumber \\
|\Psi_2^{[4]}\rangle_{(1234)}&=&1/(2\sqrt{3})
(2|0011\rangle-|0101\rangle-|0110\rangle  \nonumber \\
&-&|1001\rangle
-|1010\rangle+2|1100\rangle)_{(1234)} \nonumber \\
|\Psi_3^{[4]}\rangle_{(1234)}\! &\!=\!&\!|\Psi_1^{[4]}\rangle_{(1432)}
\label{wavefun}
\end{eqnarray}
where $|1001\rangle_{(1234)}\!=\!|1\rangle_{1}|0\rangle_{2}|0\rangle_{3}
|1\rangle_{4}$ and so on. We also compare these four-qubit DF states  
with two qubit Bell states:
$|a\rangle \!\equiv$
$(|\downarrow\downarrow\rangle $+$|\uparrow\uparrow\rangle )/\sqrt{2}$, 
$|b\rangle \!\equiv$
$(|\downarrow\downarrow\rangle$ -$|\uparrow\uparrow\rangle )/\sqrt{2}$, 
$|c\rangle \!\equiv$
$(|\downarrow\uparrow\rangle$ +$|\uparrow\downarrow\rangle )/\sqrt{2}$, 
$|d\rangle \!\equiv$
$(|\downarrow\uparrow\rangle$ -$|\uparrow\downarrow\rangle  )/\sqrt{2}$ 
depicted in Fig.~\ref{QPC} (b). 

\section{Formulation}
We show the formulation for Fig.~\ref{QPC}(a). 
The Hamiltonian for the combined qubits and the QPC for Fig.~\ref{QPC}(a) 
is written as $H = H_{\rm qb}\!+\!H_{\rm qpc}\!+\!H_{\rm int}$.  
$H_{\rm qb}$ describes the interacting four qubits:
$ 
H_{\rm qb}\!=\!\sum_{i=1}^N \!
\left(\Omega_i \sigma_{i x} \!+\!
\epsilon_i \sigma_{i z} \right)
\!+\! \!\sum_{i=1}^{N\!-\!1}
 \!J_{i,i+1} \sigma_{i z} \sigma_{i+1z}, 
$ 
where $\Omega_i$ and $\epsilon_i$ are the inter-QD tunnel coupling and
energy difference (gate bias) within each qubit.  
Here, spin operators are used instead of
annihilation operators of an electron in each qubit.  
$J_{i,i+1}$ is a coupling constant between two nearest qubits,
originating from capacitive couplings in the QD system \cite{tana0}.
$|\!\uparrow\rangle$ and $|\!\downarrow\rangle$ refer
to the two single-qubit states in which the excess charge is localized in the
upper and lower dot, respectively.  

The two serially coupled QPCs are described by
\begin{eqnarray}
H_{\rm qpc} \!\!&=&\!\! \!\! 
\sum_{\alpha=L,R \atop s\!=\!\uparrow,\downarrow}\! 
\sum_{\ i_\alpha} \!\!\left[ E_{i_\alpha} c_{i_\alpha s}^\dagger
c_{i_\alpha s} \!+\! V_{i_\alpha s} (c_{i_\alpha s}^\dagger d_{s} +
d_{s}^\dagger c_{i_\alpha s} )\right]
\nonumber \\
&+& \sum_{s\!=\!\uparrow,\downarrow} E_d d_{s}^\dagger d_{s} 
+ U d_{\uparrow}^\dagger d_{\uparrow} d_{\downarrow}^\dagger d_{\downarrow}\,.
\label{eqn:H_qpc}
\end{eqnarray}
\normalsize
Here, $c_{i_{L}s}$($c_{i_{R}s}$) is the annihilation operator of an electron
in the $i_L$th ($i_R$th) level ($i_L(i_R)=1,...,n)$ of the left (right)
electrode, $d_{s}$ is the electron annihilation operator of the island
between the QPCs, $E_{i_{L}s}$($E_{i_{R}s}$) is the energy level of electrons 
in the left (right) electrode, and $E_d$ is that of the island.  Here, we
assume only one energy level on the island between the two QPCs, with spin
degeneracy.  $V_{i_Ls}$ ($V_{i_Rs}$) is the tunneling strength of electrons
between the left (right) electrode state $i_Ls$ ($i_Rs$) and the island state.  
$U$ is the on-site Coulomb energy of double occupancy in the island. 

$H_{\rm int}$ is the capacitive interaction between the qubits and the QPC,
that induces {\it dephasing} between different eigenstates of $\sigma_{iz}$
\cite{Zanardi}.
Most importantly, it takes into account the fact 
that localized charge near the QPC
increases the energy of the system electrostatically, thus affecting the
tunnel coupling between the left and right electrodes: 
\begin{eqnarray}
H_{\rm int} &=& 
\sum_{i_L,s} \left[ \sum_{i=1}^2 \delta V_{i_L,is}\sigma_{iz} \right]
( c_{i_L s}^\dagger d_s + d_s^\dagger c_{i_L s} ) \nonumber \\
&+& 
\sum_{i_R,s} \left[ \sum_{i=3}^4 \delta V_{i_R, is}\sigma_{iz} \right]
( c_{i_R s}^\dagger d_s + d_s^\dagger c_{i_R s} ) 
\end{eqnarray}
where $\delta V_{i_\alpha is}$ ($\alpha=L,R$) 
is an effective change of the tunneling strength
between the electrodes and QPC island.  
Hereafter we neglect the spin dependence of $V_{i_\alpha}$ and 
$\delta V_{i_\alpha,i}$.
We assume that the tunneling strength of electrons weakly depends 
on the energy $V_{i_\alpha i}= V_{i\alpha}(E_{i_\alpha})$ 
and electrodes are degenerate up to the Fermi
surface $\mu_\alpha$.  
Then qubit states influence the QPC tunneling rate $\Gamma_L$ and $\Gamma_R$ 
by $\Gamma_L^{-1}\!=\!\Gamma_{1}^{-1}\!+\!\Gamma_{2}^{-1}$
and $\Gamma_R^{-1}\!=\!\Gamma_{3}^{-1}\!+\!\Gamma_{4}^{-1}$ 
through $\Gamma_{i}^{(\pm)} \equiv
2\pi \wp_{\alpha} (\mu_\alpha) |V_{i\alpha}^{(\pm)} (\mu_\alpha) |^2$ and
$\Gamma_{i}^{(\pm)'} \!
\equiv\! 2\pi \wp_{\alpha} (\mu_\alpha\!+\!U) 
|V_{i\alpha}^{(\pm)} (\mu_\alpha\!+\!U)|^2$, 
depending on the qubit state 
$\sigma_{iz}=\pm 1$ 
($V_{i\alpha}^{(\pm)}(\mu_\alpha) =
V_{i\alpha}\!(\mu_\alpha) \pm \delta V_{i\alpha}(\mu_\alpha)$ 
and $\wp_{\alpha} (\mu_\alpha)$ is the density of states of the electrodes
($\alpha\!=\!L,R$)).  
The values of $\Gamma_{i}^{(\pm)}$s are determined 
by the geometrical structure
of the system.
The strength of measurement is parameterized by 
$\Delta \Gamma_{i}$ as 
$\Gamma_{i}^{(\pm)}\!=\!\Gamma_{i0}\!\pm\Delta \Gamma_{i}$.
The measurement strength $\zeta$ is related to the tunneling rates 
as $\Gamma_i\!=\!\Gamma_0 (1\pm\zeta)$ ($\Gamma_0$ is an unit)\cite{tana2}.
We call $|\downarrow \downarrow \rangle$, $|\downarrow
\uparrow \rangle$, $|\uparrow \downarrow \rangle$, and $|\uparrow \uparrow
\rangle$ $|A\rangle \sim |D\rangle$ respectively, 
and four-qubit states are written as $|AA\rangle$, $|AB\rangle$,
..,$|DD\rangle$. For uniform two qubits, 
$\Gamma_A\!=\!\Gamma_0(1\!-\!\zeta)/2$, 
$\Gamma_B\!=\!\Gamma_C$ $\!=\!\Gamma_0(1\!-\!\zeta^2)/2$ and
$\Gamma_D\!=\!\Gamma_0(1\!+\!\zeta)/2$ with $\zeta\equiv \Delta\Gamma/\Gamma_0$.

The DM equations of four qubits and detector at zero temperature 
of Fig.~\ref{QPC}(a) are derived as in Ref.\cite{tana2} by
\small
\begin{eqnarray}
& &\frac{d \rho_{z_1z_2}^{a}}{dt}\!=\!(\!i[J_{z_2}\!-\!J_{z_1}\!]
-\![\Gamma_L^{z_1}\!+\!\Gamma_L^{z_2} ])\rho_{z_1z_2}^{a}
\nonumber \\
\!&\!-\!&\! i\!\sum_{j=1}^{N}\Omega_j 
(\rho_{g_j(z_1),z_2}^{a}\!-\!\rho_{z_1,g_j(z_2)}^{a})
\!+\! \sqrt{\Gamma_R^{z_1}\Gamma_R^{z_2}} 
(\rho_{z_1z_2}^{b\uparrow}+\rho_{z_1z_2}^{b\downarrow}),
\nonumber \\
& &\frac{d \rho_{z_1z_2}^{b_s}}{dt}\!=\!\left(\!i[J_{z_2}\!-\!J_{z_1}]\!
-\!\frac{\Gamma_L^{z_1'} \!+\!\Gamma_L^{z_2'} 
\!+\!\Gamma_R^{z_1}\!+\!\Gamma_R^{z_2}}{2}\right)
\rho_{z_1z_2}^{b_s}
\nonumber \\
\!&\!-\!&\! i\!\sum_{j=1}^N \Omega_j 
(\rho_{\!g_r(z_1),z_2}^{b_s}\!\!-\!\!\rho_{\!z_1,g_r(z_2)}^{b_s})
\!+\! \!
\sqrt{\Gamma_L^{z_1}\Gamma_L^{z_2}} \!\rho_{z_1\!z_2}^{a}\!
\!+\! \!\sqrt{\Gamma_R^{z_1'}\Gamma_R^{z_2'}}\! \rho_{z_1\!z_2}^{c},
\nonumber \\
& &\frac{d \rho_{z_1z_2}^{c}}{dt}
\!=\!(\!i[J_{z_2}\!-\!J_{z_1}]
\!-\![\Gamma_R^{z_1'}+\Gamma_R^{z_2'}] )
\rho_{z_1z_2}^{c}
\nonumber \\
& &\!-\! i\!\sum_{j=1}^N \Omega_j 
(\rho_{g_r(z_1),z_2}^{c}\!-\!\rho_{z_1,g_r(z_2)}^{c})
\!+\! \sqrt{\Gamma_L^{z_1'}\Gamma_L^{z_2'}}
(\rho_{z_1z_2}^{b\uparrow}+\rho_{z_1z_2}^{b\downarrow}),
\nonumber\\ \!\!\!\!\!\!\!\!\!\!\!\!\!\!\!\!\!\!\!\!\!
\label{eqn:dm}
\end{eqnarray}
\normalsize
where $z_1,z_2=AA,AB,...,DD$ and, 
$\rho_{z_1z_2}^{a}$, $\rho_{z_1z_2}^{b_s}$ and 
$\rho_{z_1z_2}^{c}$ are density matrix elements 
when no electron, one electron and two electrons 
exist in the QPC island, respectively. 
$J_{AA} \!=\! \sum_i^4 \epsilon_i\!+\!J_{12}\!+\!J_{23}$, 
$J_{AB} \!=\! \sum_i^3 \epsilon_i\!-\!\epsilon_4\!+\!J_{12}\!-\!J_{23}$, 
...,
$J_{DD} \!=\!-\sum_i^4\epsilon_i\!+\!J_{12}\!+\!J_{23}$.
$g_l(z_i)$ and $g_r(z_i)$ are introduced for the sake of notational 
convenience and determined by the relative positions between qubit
states as in Ref.~\cite{tana2}. 
We have 768 equations for four-qubits. 
\begin{figure}
\begin{center}
\includegraphics[width=6cm]{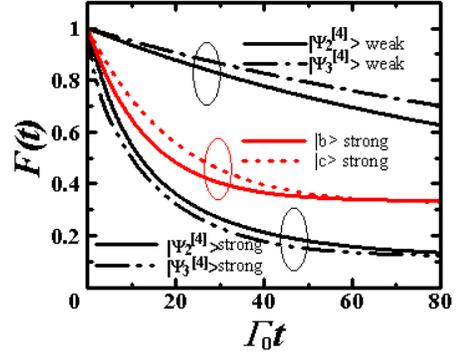}
\end{center}
\caption{Time-dependent fidelity of four-qubit states 
($|\Psi_2^{[4]}\rangle$ and $|\Psi_3^{[4]}\rangle$) and 
two-qubit non-DF states ($|b\rangle$ and $|c\rangle$). 
The 'weak' means a weak measurement case of 
$\zeta=0.2$ and the 'strong' means a strong measurement 
case of $\zeta=0.6$.
$\Omega=2\Gamma_0$, $J_{ij}=0$ $\epsilon_i=0$.
$\Gamma'_{i}=\Gamma_{i}$.
}
\label{Fstruct}
\end{figure}

To see the decoherence effect explicitly, we study 
time-dependent {\it fidelity}, 
$F(t)\!\equiv\! {\rm Tr}[\rho(0) \rho'(t)]$ 
on the rotating coordinate as 
$ 
\hat{\rho}'(t)\!=$$e^{i\sum\Omega_i'\sigma_{ix} t} \hat{\rho} (t) 
e^{-i\sum\Omega_i'\sigma_{ix} t}
$ ($\Omega_i'\!\equiv\! \sqrt{\Omega_i^2\!+\!\epsilon_i^2/4}$) to eliminate 
the bonding-antibonding coherent oscillations of free qubits
(trace is carried out over qubit states).

\section{Numerical Results}
Figure~\ref{Fstruct}
shows the effect of the local structure, that is, an island 
in the QPC detector, on the fidelity of DF states. It is seen that 
the local structure greatly degrades the fidelity of qubit states. 
In particular, the degradation is large 
when the strength of measurement increases. 
Figure~\ref{Fstruct} also shows that non-DF two-qubit states 
are better than four qubit DF states when the strength of measurement 
is large. 
This result can be understood if we consider that 
the DF states that include many qubits have a disadvantage in
that they are sensitive to local structures around them 
because they are distributed widely.
In Ref\cite{tana2},
we showed that $|b \rangle$ state is a candidate of logical 
state in the weak measurement case, and $|c\rangle$ is a candidate 
in the strong measurement case, 
by exactly solving the DM equations analytically.
Figure~\ref{Fstruct} supports the view that 
$|c \rangle$ state is better in the strong measurement case. 
\begin{figure}
\begin{center}
\includegraphics[width=7cm]{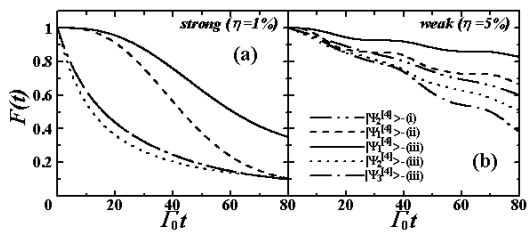}
\end{center}
\caption{Time-dependent fidelity 
of four-qubit DF states ($|\Psi_1^{[4]}\rangle$, $|\Psi_2^{[4]}\rangle$ and 
$|\Psi_3^{[4]}\rangle$)
under various fluctuations : 
(i)$\Omega_3\!=\!(1\!-\!\eta)\Omega$,
$\epsilon_3\!=\!\eta\Gamma_0$ and 
$\Gamma_3^{(\pm)}\!=\!(1\!-\!\eta)\Gamma^{(\pm)}$. 
(ii)$\Omega_2\!=\!\Omega_3\!=\!(1\!-\!\eta)\Omega$,
$\epsilon_2\!=\!\epsilon_3\!=\!\eta\Gamma_0$ and 
$\Gamma_2^{(\pm)}=\Gamma_3^{(\pm)}\!=\!(1\!-\!\eta)\Gamma^{(\pm)}$.
(iii)$\Omega_4\!=\!(1\!-\!\eta)\Omega$,
$\epsilon_4\!=\!\eta\Gamma_0$ and 
$\Gamma_4^{(\pm)}\!=\!(1\!-\!\eta)\Gamma^{(\pm)}$.
(a) $\eta$=0.01 and $\zeta$=0.6 (strong measurement), 
(b) $\eta$=0.05 and $\zeta$=0.2 (weak measurement). 
$\Omega=2\Gamma_0$, $J_{ij}=0$ $\epsilon_i=0$.
$\Gamma'_{i}=\Gamma_{i}$.
}
\label{FID}
\end{figure}

Figure~\ref{FID} shows the combined effect of 
local QPC island structure and the non-uniformities of the qubits.
In case (i), 
only 3rd qubit fluctuates 
as $\Omega_3\!\rightarrow\! \Omega_3(1\!-\!\eta)$, 
$\epsilon_3\!\rightarrow\! \epsilon_3(1\!-\!\eta)$ and
$\Gamma_3\!\rightarrow\! \Gamma_3(1\!-\!\eta)$. 
In case (ii), the 2nd and 3rd qubits fluctuate. In case (iii)
only 4th qubit fluctuates. All of these non-uniformities 
are introduced in a manner similar to that described in Ref.~\cite{tana2}.

Figure~\ref{FID} (a) shows that non-uniformity of qubit 
does not change fidelity of $|\Psi_2^{[4]}\rangle$ and $|\Psi_3^{[4]}\rangle$ 
in the strong measurement case, when compared with Fig.~\ref{Fstruct}. 
This is because the local structure has already greatly degraded the 
fidelity without non-uniformity as shown in Fig.~\ref{Fstruct}.
On the other hand, $|\Psi_1^{[4]}\rangle$, which is a product state 
of two singlet states, degrades due to the non-uniformity. 
Thus, the combination of local structure and non-uniformity 
degrades the four-qubit states even when the non-uniformity is small (1\%).

Figure~\ref{FID} (b) shows the robustness of four-qubit DF states 
depends on the distribution of non-uniformities in the qubits.
Similar to the results in Ref.~\cite{tana2}, 
the fidelities of case (iii) in  $|\Psi_2^{[4]}\rangle$ and $|\Psi_3^{[4]}\rangle$ 
are smaller than those of other distributions of non-uniformities.
For $|\Psi_1^{[4]}\rangle$, the distribution of case (ii) is 
the largest, because 
both the two singlets 
$2^{-1}(|01\rangle\!-\!|10\rangle)_{(12)}$ and 
$(|01\rangle\!-\!|10\rangle)_{(34)}$ are affected in the configuration 
of non-uniformity.

\begin{figure}
\begin{center}
\includegraphics[width=7cm]{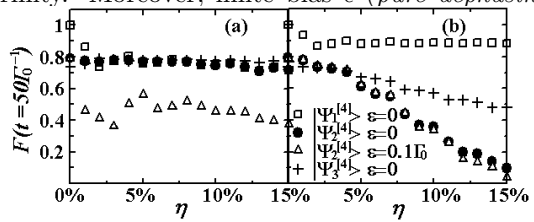}
\end{center}
\caption{Fidelities of four-qubit DF states at $t\!=\!50\Gamma_0^{-1}$ 
as a function of non-uniformity $\eta$.
(a) Non-uniformity for 2nd and 3rd qubits (case (ii)). 
(b) Non-uniformity for 4th qubit (case (iii)). 
$\Omega=2\Gamma_0$, $J_{ij}=0$ and $\zeta$=0.2.
}
\label{eta}
\end{figure}

Figures~\ref{eta} (a) and (b) show the fidelities for four-qubit DF 
states as a function of non-uniformity $\eta$ at $\Gamma_0 t=50$.
Basically, as $\eta$ increases, the fidelity decreases and 
the degradation strongly depends on the distribution 
of non-uniformity. 
Moreover, finite bias $\epsilon$ ({\it pure dephasing} region)  
reduces the fidelity more than in the zero bias cases.
Figure~\ref{eta} (b) can be compared with the Fig. 3
in Ref.\cite{tana2} where QPC detector has no island structure:
fidelity of the present island QPC detector is less 
than that of the structureless QPC in Ref.\cite{tana2}. 
In particular, even at $\eta=0$, fidelity in the present 
case degrades because the island in the QPC detector 
violates the collective decoherence environment.

\section{Conclusion}
We have solved master equations of four and two qubits 
with QPC detector, and discuss the robustness of 
DF states when there are a local structure and non-uniformities. 
We found that local structure is an obstacle to 
using DF states other than non-uniformities.
We also showed that two-qubit non-DF states are 
candidates for the logical qubits even when there 
is some local structure.



\end{document}